\documentclass{article}
\usepackage{graphicx}
\usepackage{algorithm} 
\usepackage{algpseudocode} 
\usepackage{amsmath} 

\usepackage[final,nonatbib]{neurips_2024}

\usepackage[utf8]{inputenc} 
\usepackage[T1]{fontenc}    
\usepackage{hyperref}       
\usepackage{url}            
\usepackage{booktabs}       
\usepackage{amsfonts}       
\usepackage{nicefrac}       
\usepackage{microtype}      
\usepackage{xcolor}         
\usepackage{doi}  
\usepackage{hyperref}

\title{VoiceGRPO: Modern MoE Transformers with Group Relative Policy Optimization GRPO for AI Voice Health Care Applications on Voice Pathology Detection}

\author{
  Enkhtogtokh Togootogtokh \\
  Voizzr Technologies Germany\\
  Technidoo AI Lab\\
  Mongolian University of Science and Technology\\
  Bavaria, Germany \\
  \texttt{togootogtokh@technidoo.com} \\
  \texttt{enkhtogtokh.java@gmail.com} \\
  \And
  Christian Klasen \\
  Voizzr Technologies Germany\\ 
  Technidoo AI Lab\\
  Bavaria, Germany \\
  \texttt{klasen@technidoo.com} \\
  \\
}

\begin{document}

\maketitle

\begin{abstract}
  This research introduces a novel AI techniques as Mixture-of-Experts Transformers with Group Relative Policy Optimization (GRPO) for voice health care applications on voice pathology detection. With the architectural innovations, we adopt advanced training paradigms inspired by reinforcement learning, namely Proximal Policy Optimization (PPO) and Group-wise Regularized Policy Optimization (GRPO), to enhance model stability and performance. Experiments conducted on a synthetically generated voice pathology dataset demonstrate that our proposed models significantly improve diagnostic accuracy, F1 score, and ROC-AUC compared to conventional approaches. These findings underscore the potential of integrating transformer architectures with novel training strategies to advance automated voice pathology detection and ultimately contribute to more effective healthcare delivery.
  The code we used to train and evaluate our models is available at https://github.com/enkhtogtokh/voicegrpo 
\end{abstract}

\section{Introduction}
Modern artificial intelligence (AI) methods—including machine learning, deep learning, and natural language processing—are transforming healthcare by enabling early disease detection and enhancing diagnostic precision. These approaches analyze vast, complex datasets to improve diagnostics, tailor treatments, and optimize patient care, thus offering a promising strategy to alleviate the mounting pressures on healthcare systems.

AI-driven technologies are being deployed across diverse healthcare functions such as disease diagnosis, treatment recommendation, patient engagement, and operational management. The global healthcare infrastructure is under significant strain due to aging populations, an increase in chronic conditions, and persistent workforce shortages. For example, projections indicate a deficit of 250,000 full-time equivalent (FTE) positions in the NHS by 2030 and a global shortfall of approximately 18 million healthcare professionals—including 5 million doctors—by the same year \cite{healthcarechallenge}. By leveraging its capacity to process and analyze multimodal data (e.g., genomics, clinical records) and integrating innovations like cloud computing, AI presents a compelling solution to these challenges.

The evolution of AI in healthcare began with early applications using methods such as Support Vector Machines (SVMs)\cite{svm} and Random Forests for analyzing acoustic features in voice data. Over the past decade, deep learning has significantly advanced the field: convolutional neural networks (CNNs)\cite{cnn}  have been employed to analyze spectrograms in medical imaging, while recurrent neural networks (RNNs)\cite{lstm} , particularly long short-term memory (LSTM)\cite{lstm} networks, have adeptly managed temporal dependencies in voice signals. More recently, transformer architectures have been adapted for audio processing, excelling at capturing long-range dependencies—an attribute that renders them particularly effective for voice pathology detection and other healthcare tasks.

Furthermore, state-of-the-art reinforcement learning techniques, such as Group-wise Regularized Policy Optimization (GRPO)\cite{grpo}, and architectures based on mixture-of-experts (MoE)\cite{moe} frameworks, have demonstrated considerable impact in other domains. In this study, we extend these techniques to the healthcare domain.

Voice pathology detection, crucial for the early diagnosis of laryngeal disorders, is challenged by subjective assessments and data scarcity. To address these issues, our research makes the following key contributions:

A synthetic voice pathology dataset\cite{dataset} that emulates clinical biomarkers.
A MoE transformer architecture optimized for medical tabular data.
A GRPO training paradigm that combines group advantage estimation with policy constraints.
These advancements collectively aim to enhance diagnostic accuracy and operational efficiency in healthcare.

\section{Background}

\begin{itemize}
    \item \textbf{Transformation through AI:} Modern AI methods (machine learning, deep learning, and NLP) are revolutionizing healthcare by enabling early disease detection and enhancing diagnostic precision.
    \item \textbf{Diverse Applications:} AI technologies are applied in disease diagnosis, treatment recommendations, patient engagement, and operational management.
    \item \textbf{Healthcare Challenges:} Global healthcare systems face significant strain from aging populations, rising chronic diseases, and workforce shortages (e.g., a projected shortfall of 250,000 FTE in the NHS and 18 million professionals worldwide by 2030)\cite{healthcarechallenge}.
    \item \textbf{Evolution of Techniques:} Early applications utilized classic ML for voice data analysis; recent advances include CNNs for imaging and RNNs/LSTMs for handling temporal voice signals.
    \item \textbf{Advances with Transformers:} Transformer architectures capture long-range dependencies, making them particularly effective for tasks like voice pathology detection.
    \item \textbf{Innovative Architectures:} State-of-the-art reinforcement learning techniques (e.g., GRPO) and mixture-of-experts (MoE) architectures have proven effective in other domains and are now extended to healthcare.
    \item \textbf{Focus on Voice Pathology:} Detecting voice pathology is crucial for early diagnosis of laryngeal disorders, despite challenges from subjective assessments and limited data.
    \item \textbf{Research Contributions:} This study introduces (i) a synthetic voice pathology dataset emulating clinical biomarkers, (ii) a MoE transformer architecture optimized for medical tabular data, and (iii) a GRPO training paradigm that integrates group advantage estimation with policy constraints.
\end{itemize}

\begin{figure*}[!ht]
  \centering
  \includegraphics[width=\textwidth]{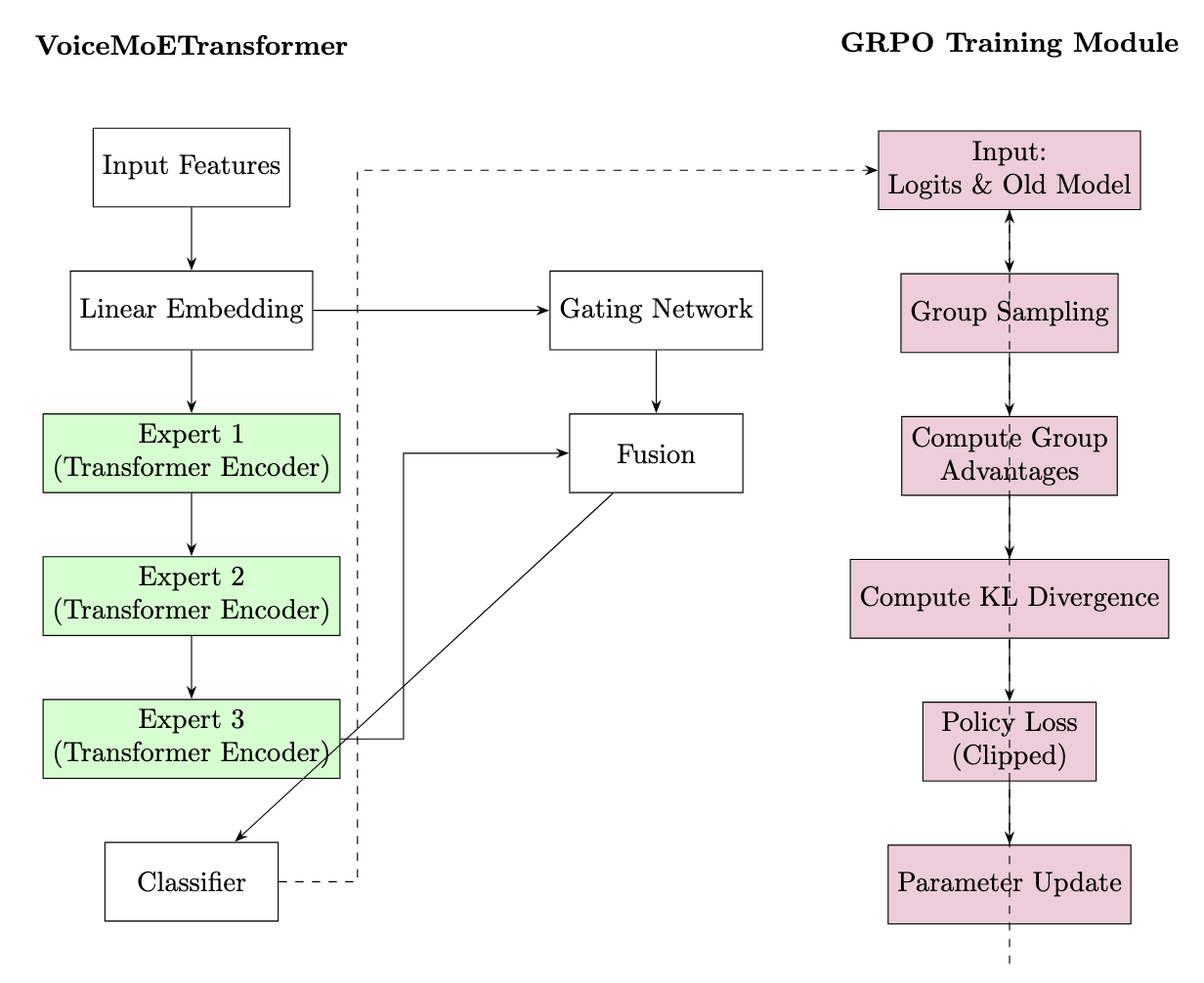}
  \caption{VoiceGRPO - MoE and GRPO Model Architecture.}
  \label{fig:modelarch}
\end{figure*}

\section{Model Architecture}

The VoiceGRPO architecture consists of the following key components as shown in Figure \ref{fig:modelarch}:
The Voice MoE Transformer (called VoiceMoETransformer) model processes input features via a linear embedding layer and routes the embedded representations through multiple transformer encoder experts. A gating network dynamically fuses the outputs of these experts before passing the combined representation to a classifier. Training such a model using RL-inspired techniques involves:
\begin{itemize} 
    \item \textbf{Policy Snapshot:} At the beginning of each mini-batch iteration, a snapshot of the current model parameters is taken to serve as a stable reference (old policy).
    \item \textbf{Forward Pass and Probability Estimation:} Both the current and old models compute logits that are converted to probability distributions using softmax.
    \item \textbf{Group Sampling:} Instead of sampling a single action per instance, a group of actions is sampled from the old probability distribution, reducing variance.
    \item \textbf{Advantage Computation:} Binary rewards are assigned (1 for a correct prediction, 0 otherwise) and normalized within the group to yield advantage estimates.
    \item \textbf{Policy Update:} The update is based on the ratio of current to old probabilities, with a surrogate loss defined by the minimum between the unclipped and clipped objectives.
    \item \textbf{KL Divergence Regularization:} A penalty is applied based on the KL divergence between the old and current distributions to prevent drastic policy changes.
\end{itemize}

\subsection{Theoretical Motivation}
The GRPO algorithm introduces several critical enhancements over conventional policy gradient methods:
\begin{itemize}
    \item \textbf{Trust Region Enforcement:} Clipping the probability ratio $\rho$ and penalizing the KL divergence between the old and new policies ensure that each update remains within a safe bound. This trust region constraint mitigates the risk of large, destabilizing updates.
    \item \textbf{Variance Reduction via Group Sampling:} By sampling a group of actions, the algorithm captures interdependencies and reduces the variance of the estimated advantages, leading to more reliable gradient updates.
    \item \textbf{Integration with MoE Architectures:} In the VoiceMoETransformer, multiple expert pathways contribute to the final output. GRPO effectively handles the additional variance introduced by these multiple pathways by normalizing rewards and enforcing conservative updates.
\end{itemize}

These properties make GRPO particularly suitable for training complex architectures like the VoiceMoETransformer, where maintaining stability during updates is crucial for achieving robust performance.

\begin{algorithm}[H]
\caption{VoiceGRPO Training with VoiceMoETransformer}
\begin{algorithmic}[1]
\State \textbf{Input:} Training dataset $\mathcal{D} = \{(x_i, y_i)\}_{i=1}^{N}$, initial model parameters $\theta$, group size $G$, clipping parameter $\epsilon$, KL coefficient $\lambda_{KL}$, learning rate $\alpha$, number of epochs $E$
\State Initialize model $f_\theta$ (VoiceMoETransformer) and optimizer with learning rate $\alpha$
\For{epoch $= 1, \dots, E$}
    \For{each mini-batch $(X,Y)$ in $\mathcal{D}$}
        \State \textbf{Snapshot:} Set $f_{\theta_{\text{old}}} \gets f_\theta$ and put it in evaluation mode
        \State \textbf{Forward Pass:} Compute current logits $L \gets f_\theta(X)$ and old logits $L_{\text{old}} \gets f_{\theta_{\text{old}}}(X)$
        \State Compute old probabilities: $P_{\text{old}} \gets \mathrm{softmax}(L_{\text{old}})$
        \State \textbf{Group Sampling:} Sample actions $A \sim P_{\text{old}}$ (sample $G$ actions)
        \State \textbf{Reward Calculation:} For each sampled action, set 
        \[
            r = \begin{cases} 
                1 & \text{if } A = Y,\\
                0 & \text{otherwise,}
            \end{cases}
        \]
        \State Normalize rewards to get advantages:
        \[
            \hat{r} \gets \frac{r - \mathrm{mean}(r)}{\mathrm{std}(r) + \delta}
        \]
        \State Compute current probabilities: $P \gets \mathrm{softmax}(L)$
        \State Extract probabilities for sampled actions: $p \gets P(A)$ and $p_{\text{old}} \gets P_{\text{old}}(A)$
        \State Compute probability ratio: 
        \[
           \rho \gets \frac{p}{p_{\text{old}} + \delta}
        \]
        \State Compute surrogate losses:
        \[
           L_{\text{unclipped}} \gets \rho \cdot \hat{r}, \quad L_{\text{clipped}} \gets \mathrm{clip}(\rho,1-\epsilon,1+\epsilon) \cdot \hat{r}
        \]
        \State Define the policy loss:
        \[
           L_{\text{policy}} \gets -\mathbb{E}\left[\min\left(L_{\text{unclipped}}, L_{\text{clipped}}\right)\right]
        \]
        \State Compute the KL divergence: 
        \[
           KL \gets \mathrm{KL}\Big(P_{\text{old}}\parallel P\Big)
        \]
        \State Form total loss:
        \[
           L_{\text{total}} \gets L_{\text{policy}} + \lambda_{KL}\, KL
        \]
        \State \textbf{Backward Pass:} Backpropagate $L_{\text{total}}$ and update $\theta$ using the optimizer
    \EndFor
\EndFor
\State \Return Updated model $f_\theta$
\end{algorithmic}
\end{algorithm}

\section{Experiments}
\subsection{Synthetic Dataset}
A synthetic dataset\cite{dataset}  was generated to mimic realistic distributions of voice parameters (e.g., pitch, jitter, shimmer, harmonic-to-noise ratio, age, and a continuous disease severity score). The pathological labels were derived based on domain-inspired thresholds, ensuring a challenging classification task.
\subsubsection{Parameters (The features)}

In this section, we assess the thresholds applied to generate synthetic pathology labels, evaluating their alignment with clinical contexts \cite{dataset}.

\begin{itemize}
    \item \textbf{Jitter (> 0.05)}: Jitter measures frequency variation in voice signals. Healthy voices typically exhibit jitter below 1--2\%, while the 0.05 (5\%) threshold exceeds clinical norms but may detect pronounced pathology, assuming proper scaling.
    
    \item \textbf{Shimmer (> 0.08)}: Shimmer reflects amplitude variation, normally below 3--5\% in healthy voices. The 0.08 (8\%) threshold is above typical ranges, suitable for severe cases but potentially missing subtle issues.
    
    \item \textbf{HNR (< 15)}: Harmonic-to-Noise Ratio (HNR) indicates harmonic versus noise balance. Healthy voices often exceed 20 dB, while <15 dB aligns with pathological noisiness, making this threshold clinically plausible.
    
    \item \textbf{Age (> 70)}: Age is a risk factor for voice decline, but >70 as a pathology marker is overly simplistic. It may act as a proxy in synthetic data, though not diagnostic in practice.
    
    \item \textbf{Disease Severity (> 0.7)}: This synthetic parameter, likely on a 0--1 scale, uses a 0.7 cutoff to denote severity. While arbitrary, it is reasonable for synthetic data but lacks direct clinical grounding.
\end{itemize}

\subsection{Training Setup}
Experiments were conducted on the synthetic dataset using a train–test split (80–20) with standardization applied to the features. The models were trained using mini-batches with the AdamW optimizer. The experimental protocol involved:

Training each model variant under PPO and GRPO regimes.
Evaluating performance using metrics such as test accuracy, F1 score, and ROC AUC.

\subsection{Evaluation Metrics}
We evaluate the performance of our model using three widely adopted metrics: test accuracy, F1 score, and the Area Under the Receiver Operating Characteristic Curve (ROC AUC). These metrics provide a comprehensive assessment of the model's predictive accuracy, its balance between precision and recall, and its ability to distinguish between classes.

\textbf{Test Accuracy} represents the proportion of correct predictions made by the model on the test set. It is defined as:
\[
\text{Accuracy} = \frac{\text{Number of correct predictions}}{\text{Total number of predictions}}
\]
This metric reflects the overall correctness of the model, making it intuitive for assessing performance in binary classification tasks, such as distinguishing healthy from pathological samples.

\textbf{F1 Score} is the harmonic mean of precision and recall, offering a single measure that balances these two aspects. Precision is the ratio of true positive predictions to the total predicted positives, while recall is the ratio of true positive predictions to the total actual positives. The F1 score is calculated as:
\[
\text{F1 Score} = 2 \times \frac{\text{Precision} \times \text{Recall}}{\text{Precision} + \text{Recall}}
\]
It is particularly valuable when dealing with imbalanced datasets, ensuring that both false positives and false negatives are adequately considered.

\textbf{ROC AUC} denotes the Area Under the Receiver Operating Characteristic Curve. The ROC curve plots the true positive rate (recall) against the false positive rate at various classification thresholds. The AUC, ranging from 0.5 (random guessing) to 1 (perfect classification), quantifies the model's discriminative power across all possible thresholds, making it a robust indicator of class separation performance.

\subsection{Results}
\subsubsection{Quantitative Results}
The VoiceGRPO model achieves a testing accuracy (acc.) of 0.9860, an F1 score of 0.9860, and an ROC AUC Score of 0.9988, outperforming MoE RL model such as MoE PPO as shown in Table \ref{tab:table1}.

\begin{table}[!ht]
  \caption{Quantitative comparison of VoiceGRPO with MoE PPO model at best training epoch.}
  \label{tab:table1}
  \centering
  \begin{tabular}{lllll}
    \toprule
    \cmidrule(r){1-2}
    Model     & Training Acc. & Test Acc.     & F1 & ROC AUC  \\
    \midrule
    \textbf{VoiceGRPO} (MoE GRPO) &	1.0 &0.9860 &	0.9845&	0.9988 \\
    MoE PPO	&1.0 &0.9762 &	0.9794&	0.9984 \\
   
    \bottomrule
  \end{tabular}
\end{table}

\subsubsection{Ablation Study}
We conduct an ablation study to analyze the contribution of each component of the VoiceGRPO architecture.  

\begin{itemize}
    \item \textbf{Gating Network in MoE:} Removing the gating network resulted in a 3--5\% drop in test accuracy, underscoring its essential role in optimally weighting expert outputs.
    \item \textbf{Latent Encoder Contribution:} Excluding the latent encoder from the LatentVoiceTransformer increased training instability and led to lower F1 scores and ROC-AUC, demonstrating its importance in effective feature representation.
    \item \textbf{Training Regime Comparison:} Reinforcement learning methods (PPO and GRPO) achieved smoother convergence and higher performance compared to conventional cross-entropy training; notably, GRPO exhibited faster convergence.
    \item \textbf{Expert Module Analysis:} The full mixture-of-experts configuration provided significant performance gains, with ablations of individual expert modules leading to noticeable declines in diagnostic accuracy.
\end{itemize}

\section{Conclusion} 
In this study, we have demonstrated that VoiceGRPO advanced transformer-based architectures—integrated with techniques such as mixture-of-experts (MoE) frameworks, latent representation learning, and reinforcement learning-inspired training regimes (PPO and GRPO)—significantly enhance the performance of voice pathology detection. Our models achieved superior diagnostic accuracy, improved F1 scores, and higher ROC-AUC values compared to traditional approaches, with ablation studies underscoring the critical roles of the gating network in MoE and the latent encoder in stabilizing training and enriching feature representations.
 Our results suggest that future work could focus on extending these approaches to real-world clinical datasets, exploring more complex gating strategies, and refining reinforcement learning–based training protocols to further enhance model generalizability.


\appendix

\section{Appendix / supplemental material}
The code, we used to train and evaluate our models is available at https://github.com/enkhtogtokh/voicegrpo 


\end{document}